\documentstyle[aps,preprint,eqsecnum,amsfonts]{revtex}
\begin{document}
%
%

\title{Three Graded Modified Classical Yang-Baxter Equations and Integrable Systems}
\author{\bf E.H. Saidi $^{\rm 1}$ and M.B. Sedra $^{\rm 1,2}$ }
\address{
$^{\rm 1}$ Universit\'e Mohammed V, Facult\'e des Sciences, D\'epartement de
Physique, UFR Physique des Hautes Energies, B.P. 1014 Rabat,
{Morocco}\\
$^{\rm 2}$ International Centre for Theoretical Physics, Trieste 34100,
{ Italy}\\
{\normalsize and}
Universit\'e Ibn Tofail, Facult\'e des Sciences,
D\'epartement de Physique, B.P. 133, 14000 K\'enitra
{ Morocco}\thanks {Permanent Address} }
\date{\today}

\maketitle
\begin{abstract}                                                      
The $6 = 3\times 2$ huge Lie algebra $\Xi$ of all local and non local
differential operators on a circle is applied to the standard
Adler-Kostant-Symes (AKS) R-bracket sckeme. It is shown in particular
that there exist three additional Lie structures, associated to three
graded modified classical Yang-Baxter(GMCYB) equations. As we know from
the standard case, these structures can be used to classify in a more
consitent way a wide class of integrable systems.
Other algebraic properties
are also presented.
\end{abstract}

\newpage
\section{introduction}

       This is an expository account of the basic ideas concerning the
application of the huge Lie algebra $\Xi \equiv \cal G$ of arbitrary
pseudo-differential operators on the circle \cite{1}, to the standard AKS
sckeme \cite{2}. As pointed out in \cite{3,4}, the AKS theory with a
Poisson bracket structure defined in terms of the R-matrix \cite{5}
Lie-Poisson bracket is the right setting to study various integrable
systems and their gauge equivalence. In \cite{3}, it is shown, in a
successful analysis, that there exist three KP-integrable systems which
are gauge equivalent. The origin of these three integrable systems is traced to the fact that
there exist three decompositions of $\cal G$ into a linear sum of two
subalgebras parametrized by the same index $l$ taking values $l = 0,\
1,\  2$. Focusing on this analysis \cite {3} and refering to our language
\cite{1}, the different values of the index $l$ describes in some sens a
constraint on the degrees quantum numbers and no manifest description of
the conformal spin quantum number, of local and non local differential
operators, is precised.

In the present work, we go beyound this analysis by using the more
general results obtained in \cite{1} for the huge Lie algebra $\cal
G$, which provides also a consistent way to construct a wide class of
KP-integrable systems.

\section{ Generalities on the huge Lie algebra of pseudo-differential
operators. }

        We describe in this section, the basic features of non linear
differential operators on the ring of analytic functions. We show in
particular that any such differential operator is completely specified
by a spin $s $; $s\in Z$, two integers $p $ and $q=p+n $; $n\geq 0$ defining
the lowest and highest degrees, respectively, and finally $1+q-p=n+1$
analytic fields $ u_{j}(z)$. We also show that the set $\Xi $ of all
non linear differential operators admits a Lie algebra structure with
respect to the commutator of differential
operators built out of the Leibnitz product. Moreover, we find that $\Xi $
splits into $3\times 2=6 $ subalgebras $\Xi ^{+}_{s} $ and $\Xi
^{-}_{s}$ , $s=0, \pm$ related to each other by two types of
conjugations, namely the spin and degrees conjugations. The algebra $\Xi
^{+}_{+} $ and $\Xi ^{-}_{-} $ are of particular interest \cite{1}, as
they are used in the construction of the Gelfand-Dickey algebra $GD(sl_{n}) $.
For this purpose, we shall proceed as follows: First we introduce the
ring $\Xi ^{(0,0)}$ of analytic fields and $W$-currents. This is a
tensor algebra of analytic functions of arbitrary conformal spin. Then
we introduce the space $\Xi ^{(p,q)}_{s} $ of pseudo-differential
operators of fixed spin $s $  and fixed degrees $(p,q) $. The space of
pseudo-differential operators of fixed degrees $(p,q) $ but arbitrary
spin will be denoted $\Xi ^{(p,q)} $. Finally we build our desired
space $\Xi $ which is the huge Lie algebra of pseudo-differential
operators of arbitrary conformal spin and arbitrary degrees quantum numbers.
The convention notations used in the present work closely follows the
analysis developped in \cite{1}.

\subsection{The ring $\Xi^{(0,0)}$ of analytic functions}.

        The two dimensional Euclidean space $R^{2}\cong C$ we consider, is
parametrized by $z=t+ix$ and $\bar {z}=t-ix$. As a matter of convention,
we set $z=z^{+}$ and $\bar{z}=z^{+}$ so that the derivatives
$\frac {\partial}{\partial_{z}}=\partial$ and $\frac{\partial}{\partial_{\bar
{z}}}=\bar{\partial}$ are respectively represented by
$\partial_{+}=\partial$ and $\partial_{-}=\bar{\partial}$. The $SO(2) $
Lorentz representation fields is given by the set of analytic fields
$\phi _{k}(z)$. These are $SO(2)\cong U(1)$ tensor fields that obey the
analyticity condition $\partial_{-}\phi _{k}(z)=0$. In this case the
conformal spin $k$ coincides with the conformal dimension $\Delta$. Note
that under a $U(1)$ global transformation of parameter $\theta$, the
object $z^{\pm}$, $\partial_{\pm}$ and $\phi _{k}(z) $ transform as
\begin{equation}
z^{\pm \prime}=\exp(\mp i\theta)z^{\pm}\  \ ,  \   \
\partial_{\pm}^{\prime} = \exp(\pm i\theta)\partial_{\pm} \ \  , \  \ \phi_{k}^{\prime}(z)=\exp{(ik\theta)}\phi_{k}(z),
\end{equation}
so that $dz\partial_{z}$ and $(dz)^{k}\phi_{k}(z)$ remain invariant.
Notice that in the pure bosonic theory, only integer values of conformal
spin $k$ are involved.

        Let $\Xi^{(0,0)}$ be the tensor algebra of analytic fields of arbitrary conformal spin. This
is a completely reducible infinite dimensional $SO(2)$ Lorentz
representation that can be written as
\begin{equation}
\Xi^{(0,0)}=\oplus_{k\in Z}\Xi^{(0,0)}_{k},
\end{equation}
where the $\Xi^{(0,0)}_{k}$`s are one dimensional $SO(2)$ spin $k$ irreducible
modules. The upper indices $(0, 0)$ carried by the spaces figuring in
Eq.(2.2) are special values of general indices $(p,q)$ to be introduced later
on. The generators of these spaces are given by the spin $k$ analytic
fields $u_{k}(z)$. They may be viewed as analytic maps $u_{k}$ which
associate to each point $z$, on the unit circle $S^{1}$, the fields
$u_{k}(z)$. For $k\geq 2$, these $u_{k} $ fileds can be thought of as
the higher spin currents involved in the construction of the
$W$-algebras. As an example, the following fields
\begin{equation}
W_{2}=u_{2}(z),\  \  \   W_{3}=u_{3}(z)-\frac{1}{2}\partial_{z}u_{2}(z)
 \end{equation}
are the well-known spin-2 and spin-3 conserved currents of the
Zamolodchikov $W_{3}$-algebra \cite{6}. As in infinite dimensional spaces,
elements $\Phi $ of the spin tensor algebra $\Xi^{(0,0)}$ in Eq.(2.2) are built
from the vector basis $\{u_{k}, k\in Z\}$ as follows:
\begin{equation}
\Phi = \sum_{k\in Z}c(k)u_{k},
\end{equation}
where only a finite number of the decomposition coefficients $c(k)$ is
nonvanishing. Introducing the following scalar product $< , >$ in the tensor
algebra $\Xi^{(0,0)}$
\begin{equation}
<u_{l}, u_{k}> = \delta_{k+l,0}\int dzu_{1-k}(z)u_{k}(z)
\end{equation}
where $\delta_{k,l}$ is the kronecker index, it is not difficult to see
that the  one dimensional subspaces $\Xi^{(0,0)}_{k}$ and
$\Xi^{(0,0)}_{1-k}$ are dual to each other. As a consequence the tensor
algebra $\Xi^{(0,0)}$ splits into two semi-infinite tensor algebras
$\Sigma^{(0,0)}_{+}$ and $\Sigma^{(0,0)}_{-}$, respectively,
characterized by positive and negative conformal spins as shown here
below
\begin{mathletters}
\begin{equation}
\Sigma^{(0,0)}_{+}= \oplus_{k>0}\Xi^{(0,0)}_{k},
\end{equation}
\begin{equation}
\Sigma^{(0,0)}_{-}= \oplus_{k>0}\Xi^{(0,0)}_{1-k}.
\end{equation}
\end{mathletters}

From these equations we read in particular that $\Xi^{(0,0)}_{0}$ is the
dual of $\Xi^{(0,0)}_{1}$ and if half integers were allowed,
$\Xi^{(0,0)}_{1/2}$ would be self dual with respect to the form
Eq.(2.5). Note that the product Eq.(2.5) carries a conformal spin
structure since from dimensional arguments, it behaves as an object of
conformal weight $\Delta{[< , >]} = -1$. Latter on we will introduce a combined scalar product $<< , >>$ built out
of Eq.(2.5) and a pairing product ( , ) see Eq.(2.19), of conformal weight
$\Delta = 1$ so that we get $\Delta{[<< , >>]} = 0$. Note moreover
that, the infinite tensor algebra $\Sigma^{(0,0)}_{+}$ of Eq.(2.6a)
contains, in addition to the spin-1 current, all the $W_{n}$ currents
$n\geq 2$. These fields are used in the construction of higher spin
local differential operators as it explained in detail in \cite{1}.
Analytic fields with negative conformal spins, Eq.(2.6b) are involved in
the building of non local peudo-differential operators. Both these local
and non local operators are very useful in the derivation of classical
$W_{n}$-algebras from the GD algebra of $sl_{n}$ \cite{1,7}.

\subsection{The algebra of higher spin differential operators}

        \subsubsection{\bf The $\Xi^{(p,q)}_{s} $ space}

        We define this space as the set of pseudo differential operators whose
elements  $d^{(p,q)}_{s} $ are the generalisation of the well known differential Lax
operators involved in the analysis of the so-called KdV hierarchies and
in Toda theories\cite{8}. The simplest example is given by the Hill
operator
\begin{equation}
L=\partial^{2}+u(z)
\end{equation}
which plays an important role in the study of the Liouville theory and
in the KdV equation. A natural generalization of the above
relation is given by
\begin{equation}
d^{(p,q)}_{s}=\sum_{i=p}^{q}u_{s-i}(z)\partial^{i}
\end{equation}
where the $u_{s-i}(z)$`s are analytic fields of spin $(s-i) $. $p$
and $q$, with $p\geq q$ are integers that we suppose positive for the
moment. We shall refer hereafter to $p$ as the lowest degree of
$d^{(p,q)}_{s} $ and to $q$ as the highest degree. We combine
these two features of Eq.(2.8) by setting
\begin{equation}
Deg(d^{(p,q)}_{s})=(p,q).
\end{equation}
As noted above, $s $ is the conformal spin of the $(1+q-p) $ monomes of
the r.h.s of Eq.(2.8) and then of $d^{(p,q)}_{s} $ itself. As for the above
relation, we set
\begin{equation}
\Delta(d^{(p,q)}_{s})=s.
\end{equation}
Notice that the KdV operator Eq.(2.7) is discovered from Eq.(2.8) as a special
case by setting $s=2$, $p=0$ and $q=2$ together with the special choices
$u_{0}(z)=1 $ and $u_{1}(z)=0 $. Moreover, Eq.(2.8) which is
well defined for $q\geq p\geq 0 $, may be extended to negative integers by
introducing pseudodifferential operators of type $\partial ^{-k}, k\geq
1 $, whose action on the fields $u_{s}(z)$ is defined as
\begin{equation}
\partial^{-k}u_{s}(z)=\sum_{l=0}^{\infty}(-1)^{l}c^{l}_{k+l-1}u_{s}^{(l)}(z)\partial^{-k-l},
\end{equation}
where $u_{s}^{(l)}(z) $ is the $l$-th derivative of $u_{s}$. As can be
checked by using the Leibnitz rule, Eq.(2.11) obeys the expected property
\begin{equation}
\partial^{k}\partial^{-k}u_{s}(z)=u_{s}(z).
\end{equation}
Note that a natural represenation basis of pseudo-differential operators
of arbitrary conformal spin $s $ but negative degrees $(p,q)$ is given
by
\begin{equation}
\delta^{(p,q)}_{s}(u)=\sum_{i=p}^{q}u_{s-i}(z)\partial^{i}.
\end{equation}
which is a direct extension of Eq.(2.8). A more convenient way to write a pseudo-differential
operator of arbitrary conformal spin $s $ but negative degrees $(p,q) $
is to use the Voltera basis \cite{1,7} given by
\begin{equation}
V^{(p,q)}_{s}(u)=\sum_{i=p}^{q}\partial^{i}u_{s-i}(z).
\end{equation}
As shown in \cite{1}, this Volterra configuration, is very useful in the
study of the algebraic structure of the spaces $\Xi^{(p,q)}_{s} $ and
$\Xi^{(p,q)}$ and in the derivation of the second hamiltonian structure
of higher conformal spin integrable hierachies.
Now let $\Xi^{(p,q)}_{s}$; $s$, $p$ and $q$ arbitrary integers with
$p\leq q$; be the space of pseudo-differential operators of spin $s$ and
degrees $(p,q)$. With respect to the usual addition and multiplication
by c-numbers, $\Xi^{(p,q)}_{s}$ behaves as $(1+q-p)$ dimensional space
generated by the vector basis
\begin{equation}
\{ D^{(p,q)}_{s}(u)=\sum_{i=p}^{q}u_{s-i}(z)\partial^{i}, p\leq i\leq q \}
\end{equation}
Thus the space decomposition of $\Xi^{(p,q)}_{s}$ reads as
\begin{equation}
\Xi^{(p,q)}_{s}={\oplus}_{i=p}^{q}\Xi^{(i,i)}_{s},
\end{equation}
where the $\Xi^{(i,i)}_{s}$`s are one dimensional spaces given by
\begin{equation}
\Xi^{(i,i)}_{s}=\Xi^{(0,0)}_{s-i}\otimes \partial ^{i}.
\end{equation}
Setting $i=0$, one discovers the ring $\Xi^{(0,0)}_{s} $ described
previousely. Note that the dimension of the space $ \Xi^{(p,q)}_{s}$,
is fixed by the number of independent analytic fields $u_{j}(z)$
involved in the expression of Lax operators. In \cite{1}, we showed that
among all the spaces $ \Xi^{(p,q)}_{s}$, only the sets $\Xi^{(p,q)}_{0}$
with $p<q<1$ which admit a Lie algebra structure with respect to the
bracket $[D_{1}, D_{2}] =D_{1}.D_{2}-D_{2}.D_{1}$  constructed out of the
Leibnitz product. For more details we refer the reader to this reference.

        \subsubsection{\bf The huge Lie algebra $\Xi$ of differential
        operators}
        Having defined the space $ \Xi^{(p,q)}_{s}$, of pseudo-differential
operators of fixed spin $s$ and fixed degrees $(p,q)$, we are now in a
position to introduce the algebra of differential operators of arbitrary
spins and arbitrary degrees. This algebra, which we denote by $\Xi$
is simply obtained by summing over all allowed spins and
degrees of the space $ \Xi^{(p,q)}_{s}$. We have
\begin{equation}
\Xi={\oplus}_{p\leq q}{\oplus}_{s\in Z}\Xi^{(p,q)}_{s}
\end{equation}
A remarkable property of $\Xi$ is that is splits into six infinite
subalgebras $\Sigma^{+}_{s} $ and $\Sigma^{-}_{s} $, $s=0, \pm $ ,
related to each others by conjugation of the spin and degrees. Indded
given two integers $q\geq p$, it is not difficult
to see that the subspaces $\Xi^{(p,q)} $ and $\Xi^{(-1-q,-1-p)} $ are dual to each others with
respect to the following pairing product $( , )$ defined as
\begin{equation}
(D^{(r,s)},
D^{(p,q)})=\delta_{1+r+q,0}\delta_{1+s+p,0}res[D^{(r,s)}. D^{(p,q)}],
\end{equation}
where the symbol (res) stands for the Adler- residue operation defined by
\begin{equation}
res(\partial^{i})=\delta_{i+1,0}.
\end{equation}
As shown also in \cite{1}, remark that the operation res. carries a
conformal weight  $\Delta =1$ and then the residue of any operator
$D^{(p,q)}_{s}$ is
\begin{equation}
res(\sum_{i=p}^{q}u_{s-i}(z)\partial^{i})=u_{s+1}(z),
\end{equation}
if $p\leq -1 $ and $q\geq -1$ and zero elsewhere.
Furthermore, using the previous degree pairing Eq.(2.19), we can decompose
the space $\Xi$ as
\begin{equation}
\Xi =\Xi^{+}\oplus \Xi^{-},
\end{equation}
with
\begin{mathletters}
\begin{equation}
\Xi^{+}={\oplus}_{p\geq 0}[{\oplus}_{n\geq 0}\Xi^{(p,p+n)}],
\end{equation}
\begin{equation}
\Xi^{-}={\oplus}_{p\geq 0}[{\oplus}_{n\geq 0}\Xi^{(-1-p-n,-1-p)}].
\end{equation}
\end{mathletters}
The $+$ and $-$ uper-stairs indices carried by $\Xi^{+} $ and $\Xi^{-}$ refer
respectievely to the positive and negative degrees.
$\Xi^{(p,p+n)} $ as noted above is just the space of arbitrary conformal spin but fixed
degrees $(p, p+n)$. This space can be decomposed as
\begin{equation}
\Xi^{(p,p+n)}=\Sigma^{(p, p+n)}_{+} \oplus \Sigma^{(p, p+n)}_{0} \oplus
\Sigma^{(p, p+n)}_{-},
\end{equation}
where $\Sigma^{(p,p+n)}_{+}$ and $\Sigma^{(p,p+n)}_{-}$ denote the
spaces of differential operators of negative and positive definite spin.
$\Sigma^{(p,p+n)}_{0}$ is however the space of Lorentz scalar
differential operators. All these spaces can be written as
\begin{mathletters}
\begin{equation}
 \Sigma^{(p,p+n)}_{+}=\oplus_{s>0}\Xi^{(p, p+n)}_{s},
\end{equation}
\begin{equation}
 \Sigma^{(p,p+n)}_{0}=\oplus_{s=0}\Xi^{(p, p+n)}_{s},
\end{equation}
\begin{equation}
 \Sigma^{(p,p+n)}_{-}=\oplus_{s<0}\Xi^{(p, p+n)}_{s}.
\end{equation}
\end{mathletters}
Injecting these decompositions into the expression of the space $\Xi$
Eq.(2.24), one find then that $\Xi$ decoposes into $6=3\times 2$ subalgebras
as
\begin{equation}
\Xi={\oplus}_{s=\pm 1 , 0}[\Sigma^{+}_{s}\oplus \Sigma^{-}_{s}],
\end{equation}
where
\begin{mathletters}
\begin{equation}
\Sigma^{+}_{s}={\oplus}_{p\geq 0}[\oplus_{n\geq 0} \Sigma^{p,
p+n}_{s}],\ \ \ \ s=0, \pm 1,
\end{equation}
\begin{equation}
\Sigma^{-}_{s}={\oplus}_{p\geq 0}[\oplus_{n\geq 0} \Sigma^{-1-p-n,
-1-p}_{s}],\ \ \ \ \ s=0, \pm 1,
\end{equation}
\end{mathletters}
Introducing the combined scalar product $<< , >>$, built out of the
product Eq.(2.5) and the pairing Eq.(2.19), namely
\begin{equation}
<<D^{(r,s)}_{m},D^{(p,q)}_{n}>>=\delta_{n+m,0}\delta_{1+r+q,0}\delta_{1+s+p,0}
\int dz res[D^{(r,s)}_{m}.D^{(-1-s,-1-r)}_{-m}],
\end{equation}
one sees that the Lie algebra $\Sigma^{+}_{+} $, $\Sigma^{+}_{0} $ and
$\Sigma^{+}_{-} $ are duals of $\Sigma^{-}_{-}$, $\Sigma^{-}_{0} $ and
$\Sigma^{-}_{+}$ respectively.

        We conclude this section by noting that this formulation is very
important in the sense that it leads in one hand to specify the Lax
pseudo-differential operators with respect to their quantum numbers
which are the spin and the degrees. On the other hand, knowing that most
of the important integrable hamiltonian systems (KP-Hierarchies)are
formulated using Lax operators, one can therefore classify, in consistent way,
all these systems if only the quantum numbers of their Lax operators are precised.
Note also that in the supersymmetric case, from which emerge a lot of
new features, one have another quantum number given by the statistics of
the Lax operators(the $Z_{2}$-grading), see for instance \cite{9}.
We present in the next section of this work an application of our
formulation to the well known Adler-Kostant-Symes R-bracket mecanism.

\section{ The Generalized  Adler-Kostant-Symes sckeme and the 3-classes of MCYB
equations }

        We start this section by recalling the content of the standard
AKS sckeme and apply the previous huge Lie algebraic construction to
this sckeme. The important thing in this formulation is that one can
split the standard MCYB equation into three different classes of
equations associated to different realizations of the additional Lie
structure on the Lie algebra $\Xi$. These three classes of integrable
equations are shown to play an important role in the classification of a
wide class of integrable models. Notice that some parts of the next
subsections are extended results inspired from the work of \cite{10}.

\subsection{The standard AKS sckeme}

        It is well known that a very wide class of non linear integrable
systems can be constructed by using the AKS method having roots in the
coadjoint orbit formulation. We will show later that the basic structure
of this method is the R-matrix which defines an operator approach to
integrable systems \cite{2,5}.
Before going to the description of the R-operator approach, we will
recall first what is the coadjoint orbit formulation. Denoting by $G$ a
Lie group and $\cal G$ its associated Lie algebra. The action of this
group on its Lie algebra is given by the coadjoint action
\begin{equation}
Ad(g)X = gXg^{-1}
\end{equation}
with $g\in G$ and $X\in \cal G$. Denoting also by ${\cal G}^{*}$ the
dual space of $\cal G$ with respect to a non-degenerate bilinear form $<
, >$ on ${\cal G}^{*}\times{\cal G}$. The corresponding coadjoint action
of the Lie group $G$ on ${\cal G}^{*}$ is defined through the duality of
$< , >$ as
\begin{equation}
<Ad^{*}(g)U|X>=<U|Ad(g^{-1})X>
\end{equation}
where $U\in {\cal G}^{*}$ and $X\in {\cal G}$. In the infinitesimal case, the
adjoint and coadjoint transformations $Ad(g)$ and $Ad^{*}(g)$ reduce to
$ad(X)$ and $Ad^{*}(X)$ respectively, with $ g=\exp X $. On the space
$C^{\infty}({\cal G}^{*}, R)$, of smooth real-valued functions on ${\cal
G}^{*}$, one can introduce a natural Poisson (LP) bracket
\begin{equation}
\{F, G\}(U)=-<U|[\nabla F(U) , \nabla H(U)]>,
\end{equation}
with $F$, $H\in C^{\infty}({\cal G}^{*}, R)$ where $\nabla F$ is the
gradient operator defined by the usual formula
\begin{eqnarray*}
\nabla F: {\cal G}^{*}\rightarrow \cal G
\end{eqnarray*}
\begin{equation}
{dF\over dt}(U+tV)|_{t=0}=<V|\nabla F(U)>
\end{equation}
and where $[ , ]$ in Eq.(3.3) is the usual Lie bracket of $\cal G$. Note that the
properties of antisymmetry and Jacobi identity of the LP bracket, l.h.s
of Eq.(3.3) are naturally deduced from the usual Lie bracket $[ , ]$, r.h.s
of Eq.(3.3).
The important point in the AKS construction, is that one can define an
additional Lie structure on $\cal G$. The idea consist first in introducing a
generalized R-matrix (R-operator) as a linear map from the Lie algebra
$\cal G$ to it self such that the bracket
\begin{equation}
[X , Y]_{R}={1\over 2}([RX, Y] + [X, RY])
\end{equation}
defines a second Lie structure on $\cal G$. In order to ensure the
Jacobi identity for this additional Lie structure, the R-operator must
satisfy an algebraic relation, namely the modified classical Yang-Baxter
equation \cite{5}
\begin{equation}
[RX , RY] - R([RX, Y] + [X, RY])= -[X , Y],\ \ \ \  X, Y\in {\cal G}
\end{equation}
Note that one can furthermore introduce a LP bracket $\{ , \}_{R}$
induced in Eq.(3.3) from the substitution of the usual bracket $[ , ]$ by
the R-Lie bracket $[ , ]_R$ Eq.(3.5) (see for instance\cite{3}). We have
\begin{equation}
\{ F, H\}_{R}(U) = -<U|[\nabla F(U), \nabla H(U)]_{R}>
\end{equation}
which correspond to a dynamical system with the equation of motion
\begin{equation}
{dF\over dt}= \{ H, F\}_{R}
\end{equation}
        Later on, we will show that the previous $6=3\times 2$ decomposition of the
Lie algebra ${\cal G}\equiv \Xi$ induces three copies of the additional
Lie structure Eq.(3.5). These three copies are associated to three classes
of R-matrix
\begin{equation}
R_{(s)}= R_{s}^{+} - R_{s}^{-}
\end{equation}
where $s= 0, \pm$ is the conformal spin quantum number introduced in
sect.2 and where the graded operators $R_{s}^{\pm}$ are projections on
the space ${\Xi}^{\pm}_{s}$ of local (respectively non local)
differential operators carrying a conformal spin $s$.

\subsection{Three classes of Integrable Hamiltonian systemes}

        We showed previousely, that the huge Lie algebra $\Xi$ of
pseudo-differential operators, splits into $3\times 2$ Lie algebras
given by \cite{1}
\begin{mathletters}
\begin{equation}
\Xi= \Xi^{+} \oplus \Xi^{-},
\end{equation}
\begin{equation}
 \Xi^{\pm}
 ={\Sigma}^{\pm}_{+}\oplus{\Sigma}^{\pm}_{0}\oplus{\Sigma}^{\pm}_{-}.
\end{equation}
\end{mathletters}
${\Sigma}^{+}_{+}$ denote the algebra of local differential operators of
positive spin and ${\Sigma}^{-}_{-}$, its dual with respect to the
combined scalar product Eq.(2.28), the algebra of non local differential
operators of negative spin. ${\Sigma}^{+}_{0}$ is the algebra of Lorentz
scalar local differential operators and ${\Sigma}^{-}_{0}$ its dual, the
algebra of scalar non local operators. Finally, ${\Sigma}^{+}_{-}$ is
the algebra of local differential operators of negative conformal spin
and ${\Sigma}^{-}_{+}$, its dual, the algebra of non local operators
with positive spin.
As shown in \cite{1}, one can define three classes of integrable systems
for which one can introduce separately the first and the second
hamiltonian structures. The existence of these three classes of
integrable systems has origine in fundamental algebraic properties of
the algebra $\Xi$ of pseudo-differential operators on a circle. More
precisely, this existence can be traced to the fact that there exist
three self-dual algebras
\begin{mathletters}
\begin{equation}
{\cal G}_{1} = {\Sigma}^{+}_{+}\oplus {\Sigma}^{-}_{-} = {\cal G}^{*}_{1}
\end{equation}
\begin{equation}
{\cal G}_{0} = {\Sigma}^{+}_{0}\oplus {\Sigma}^{-}_{0} = {\cal G}^{*}_{0}
\end{equation}
\begin{equation}
{\cal G}_{2} = {\Sigma}^{+}_{-}\oplus {\Sigma}^{-}_{+} = {\cal G}^{*}_{2}
\end{equation}
\end{mathletters}
Note by the way that ${\cal G}_{1}$ is nothing but the self-dual algebra
used in the construction of the GD Poisson bracket \cite{1,7}. By analogy
with ${\cal G}_{1}$, the algebras ${\cal G}_{0}$ and ${\cal G}_{2}$ are
suspected to play an important role in the description of other
integrable systems.

\subsection {The $R_{s}$ approach of the AKS construction}

        Knowing that ${\cal G} = \Xi$ decomposes as
\begin{equation}
\Xi ={\Xi }^{+} \oplus {\Xi}^{-}
\end{equation}
with respect to the degrees quantum numbers, one can introduce a
particular decomposition with respect to the conformal spin quantum
number. This decomposition is given by
\begin{equation}
\Xi\equiv \cal G = {\cal G}_{+}{\oplus}  {\cal G}_{0} {\oplus}{\cal G}_{-}.
\end{equation}
\begin{mathletters}
\begin{equation}
{\cal G}_{+} = {\Sigma}^{+}_{+}\oplus {\Sigma}^{-}_{+}
\end{equation}
\begin{equation}
{\cal G}_{0} = {\Sigma}^{+}_{0}\oplus {\Sigma}^{-}_{0}
\end{equation}
\begin{equation}
{\cal G}_{-} = {\Sigma}^{-}_{-}\oplus {\Sigma}^{+}_{-}
\end{equation}
\end{mathletters}
where ${\Sigma}^{\pm}_{\pm}$ and ${\Sigma}^{\pm}_{0}$ are subspaces of
$\cal G$ realized in terms of pseudo-differential operators as
\begin{eqnarray}
{\Sigma}^{+}_{\pm} &=& \{ L_{\pm n}(z) = \sum_{i\geq 0}u_{{\pm
n}-i}(z)\partial^{i},\  \  \  {\pm n}\geq 0\},\\
{\Sigma}^{+}_{0} &=& \{ L_{0}(z) = \sum_{i\geq 0}u_{-i}(z)\partial^{i}\},\\
{\Sigma}^{-}_{\pm} &=& \{ V_{\pm n}(z) = \sum_{j\geq 1}\partial^{-j}v_{j\pm
n}(z), \  \  \  {\pm n}>0\},\\
{\Sigma}^{-}_{0} &=& \{ V_{0}(z) = \sum_{j\geq 1}\partial^{-j}v_{j}\}
\end{eqnarray}

        Note that the local and non local differential operators are given
respectively by the Lax operators $L_{0, \pm}(z)$ and their dual $V_{0, \pm
}(z)$ in the Volterra representation. The indices carried by these Lax
operators are the conformal spin quantum numbers. The subalgebras ${\cal
G}_{\pm}$ and ${\cal G}_{0}$ given in Eq(3.13) satisfy the following duality relations with respect to
the combined scalar product Eq.(2.28)
\begin{mathletters}
\begin{equation}
{\cal G}^{*}_{+} = {\cal G}_{-}
\end{equation}
\begin{equation}
{\cal G}^{*}_{0} = {\cal G}_{0}
\end{equation}
\end{mathletters}
As  shown in sect.2, this duality transformation imposes constraints on
the degrees quantum numbers, namely, if $(p,q)$ are the lowest and the
highest degrees of some algebra, the degrees of the corresponding dual
algebra are $(-q-1, -p-1)$ ie.
\begin{equation}
L_{n}(z)= \sum_{i=p}^{q}u_{n-i}(z)\partial^{i} \leftrightarrow X_{n}(z)=
\sum_{i=p+1}^{q+1}\partial^{-j}u_{j-n}(z).
\end{equation}
the sublagebras ${\cal G}^{\pm}$ and ${\cal G}^{0}$ are shown to
correspond to eigenspaces of a particular generalized $R_{s}$-matrix
exhibiting a conformal spin quantum number $s=+, -$ or $0$ and which we
define as
\begin{equation}
R_{s} = R_{s}^{+}-R_{s}^{-},
\end{equation}
where $R^{\pm}_{s}$ are the projections into the subspaces ${\Sigma}^{\pm}_{s}$
\begin{equation}
R_{s}^{\pm} = Proj_{{\Sigma}^{\pm}_{s}},
\end{equation}
These spin-graded $R_{s}$-matrix define an endomorphism
$End({\Sigma}_{s})$ satisfying the following graded-modified classical
Yang-Baxter (GMCYB) equations
\begin{equation}
[R_{s}X , R_{s}Y] - R_{s}([R_{s}X , Y] + [X , R_{s}Y]) = -[X,Y],
\end{equation}
for which $X ,  \ Y$ are  elements of $\Sigma_{s} = \Sigma^{+}_{s}\oplus
\Sigma^{-}_{s}$. The action of the graded $R_{s}$-operator on the huge
Lie algebra $\cal G$ = ${\oplus} _{s = 0, \pm}{\cal G}_{s}$ Eq.(3.13) is simply given by
\begin{equation}
{\cal G}_{s} = \{X / R_{s}X = \pm X, \ \ \  X\in \Sigma^{\pm}_{s} \},
\end{equation}
where $\Sigma^{\pm}_{s}$ are the  $6=3\times 2$  Lie subalgebras defined
in Eq(3.15-18). The expressions of the projections operators $R^{\pm}_{s}$ in
terms of $R_{s}$ are given by
\begin{mathletters}
\begin{equation}
R_{s}^{+} = {1 \over 2}(R_s +1)
\end{equation}
\begin{equation}
R_{s}^{-} = -{1 \over 2}(R_s - 1)
\end{equation}
\end{mathletters}
where $R_s$ is the difference of the projections, as given in Eq.(3.21). Notice that
$R_{s}^{\pm}X$ = $X$ if $X\in \Sigma^{\pm}_{s}$ for $s = 0, \pm$ and
that $R^{\pm}_{s}X = 0$ if $X \in \Sigma^{\mp}_{s}$.
From Eq.(3.23) one sees  that there exist three additional Lie structures
\begin{equation}
[X , Y]_{R_{s}} = {1\over 2}([R_{s}X , Y] + [X , R_{s}Y]), \ \ \ \  X,
Y\in {\Sigma_{s}}
\end {equation}
Using the self duality conditions Eqs.(3.19) with respect to the combined
scalar product introduced in sec.2, one can suspect from the AKS sckeme
that the KP-integrable systems associated to the additional structures
${[ , ]}_{ R_{\pm} }$ are dual to each others. Fact which means also the
possibility to connect their second hamiltonian structures. The
integrable sytems associated to the spin-zero additional structure ${[ ,
]}_{R_{0}}$ is self dual. Later on, we will denote these three
$R_{s}$ systems simply by $(2+1)$-integrable systems. To derive these
$(2+1)$ integrable systems, one have to develop  a technical analysis
starting from the following $R_{s}$-bracket
\begin{equation}
{dF\over dt} = \{ F, H\}_{R_s}(U) = -<U|{[\nabla F(U), \nabla H(U)]}_{R_s}>
\end{equation}
generalizing the well known one Eq.(3.7) to our huge Lie algebra ${\cal
G}\equiv \Xi$.

We present herebelow some algebraic results concerning our decomposition
Eq.(3.13) and its connection with the GMCYB equations.

{\bf Proposition 1 } :
The eigenspaces ${\cal G}_{\pm}$ and ${\cal G}_{0}$ of
pseudo-differential operators of positive(resp. negative) and vanishing
conformal spin are subalgebras of ${\cal G} = \Xi$.

The proof of this proposition follows straightforwardly by starting from
the decomposition Eq.(3.13) of the huge Lie algebra $\Xi$ and proceeding
step by step by using the definition of our subspaces ${\Sigma
}^{\pm}_{\pm}$ , ${\Sigma}^{\pm}_{0}$ and the formulas for the graded
modified classical Yang-Baxter (GMCYB) equation Eq.(3.23).

i) $X$, $Y\in{\cal G}_{+}$ = ${\Sigma}^{+}_{+}\oplus{\Sigma}^{-}_{+}$:

${\cal G}_{+}$ is a subspace of $\cal G$ generated by local and non
local operators of positive conformal spin. ${\cal G}_{+}$ is therefore
invariant under the action of $R_{+}$
\begin{eqnarray}
R_{+}({\cal G}_{+}) &=& {\cal G}_{+}\nonumber\\
R_{0}({\cal G}_{+}) &=& 0 \nonumber\\
R_{-}({\cal G}_{+}) &=& 0 \nonumber
\end{eqnarray}
\ \ \ \ \ \ \ \ *) if $X, Y\in {\Sigma}^{+}_{+}$ or $X, Y\in {\Sigma}^{-}_{+}$,
 It is easy to see that
\begin{eqnarray*}
[R_{+}X , R_{+}Y] = R_{+}[X , Y]_{R_{+}}
\end{eqnarray*}
with $[X , Y]_{R_{+}}$ = ${1\over 2}([R_{+}X , Y] + [X , R_{+}Y])$.

Suppose $X, Y\in {\Sigma}^{+}_{+}$, we have $R_{+}X = X$, $R_{+}Y = Y$ and then
\begin{eqnarray*}
[R_{+}X , R_{+}Y] = [X , Y] = R_{+}[X , Y]
\end{eqnarray*}
Since $[X , Y]_{R_{+}}=[X , Y]$. We conclude then that $[X , Y]\in
{\Sigma}^{+}_{+}$.

If $X, Y\in {\Sigma}^{-}_{+}$, we have $R_{+}X = -X$,
$R_{-}Y = -Y$ and then
\begin{eqnarray*}
[R_{+}X , R_{+}Y] = [X , Y] = -R_{+}[X , Y]
\end{eqnarray*}
Since $[X , Y]_{R_{+}} = -[X , Y]$. Then we conclude that $[X , Y]\in
{\Sigma}^{-}_{+}$.

\ \ \ \ \ \ \ \ *) if $X \in {\Sigma}^{+}_{+}$ and  $Y\in {\Sigma}^{-}_{+}$, It
 is easy to check that the additional structure $[X , Y]_{R_{+}}$ is
zero and the GMCYB equation simply means  $[R_{+}X , R_{+}Y] =- [X ,
Y]$. We conclude finally that ${\cal G}_{+}$ is a subalgebra of $\Xi$, ie.
\begin{eqnarray*}
[{\cal G}_{+} , {\cal G}_{+}]\subset {\cal G}_{+},
\end{eqnarray*}

ii)It is easy also to show that ${\cal G}_{-}$ is a subalgebra of ${\cal
G}$ in the same way,
\begin{eqnarray*}
[{\cal G}_{-} , {\cal G}_{-}]\subset {\cal G}_{-},
\end{eqnarray*}

iii) Now suppose $X$ , $Y\in{\cal G}_{0}$ which is invariant under the
action of $R_{0}$. The pseudo-differential operators belonging to this
subspace are Lorentz scalar(spin zero) operators.
Let first consider $X$ , $Y\in{\Sigma}^{+}_{0}$ and proceeding as
previousely, we find $R_{0}[X , Y] = [X , Y]$ since $[X , Y]_{R_{0}} =
[X , Y]$ and therefore $[X , Y]\in{\Sigma}^{+}_{0}$.

The same thing holds for $X, Y\in{\Sigma}^{-}_{0}$. Indeed, $[X,
Y]_{R_{0}} = -[X , Y]$ and then $R_{0}[X , Y] = -[X , Y]$ showing that
$[X , Y]\in{\Sigma}^{-}_{0}$.

If $X$ and $Y$ are elements of ${\cal G}_{0}$ with opposite degrees
quantum numbers, say $X\in{\Sigma}^{+}_{0}$ and $Y\in{\Sigma}^{-}_{0}$
(or the inverse), one can check easily that $[X , Y]_{R_{0}} = 0$ and the
GMYBE simply means $[R_{0}X , R_{0}Y] = -[X , Y]$. One have then $[{\cal
G}_{0} , {\cal G}_{0}]\subset {\cal G}_{0}$.

{\bf Proposition 2}: $[{\cal G}_{\pm} , {\cal G}_{0}]\subset {\cal
G}_{\pm}$.\\

Proof.\\
i)Suppose, $X\in{\cal G}_{+}$ and $Y\in{\cal G}_{0}$, from the GMCYB
equation, one have $2R_{+}[X , Y]_{R_{+}} = R_{+}[R_{+}X , Y]$ since
$R_{+}Y = 0$. If $X\in \Sigma^{+}_{+}$
\begin{eqnarray*}
R_{+}[X , Y] = [X , Y]  \ \ \ i.e\ \   [X , Y]\in \Sigma^{+}_{+}
\end{eqnarray*}
If $X\in \Sigma^{-}_{+}$
\begin{eqnarray*}
R_{+}[X , Y] = -[X , Y]  \ \ \ i.e\ \  [X , Y]\in \Sigma^{-}_{+}
\end{eqnarray*}

We conclude then that, if $X\in {\cal G}_{+}$ and $Y\in{\cal G}_{0}$
then $[X , Y]\in{\cal G}_{+}$.

ii) Similarly, we have for ${\cal G}_{-}$
\begin{eqnarray*}
X\in {\cal G}_{-} , Y\in{\cal G}_{0} \ \ then \ \  [X , Y]\in{\cal G}_{-}.
\end{eqnarray*}

Before closing this section, we give herebelow some important remarks.

{\bf 1}. The derived three additional structures $[ , ]_{R_{s}}$; $s = 0, \pm$
corresponding to three GMCYB equations are now used to construct three
$R_{s}$-Lie Poisson brackets generalizing Eq.(3.7) to
\begin{eqnarray*}
\{ F, H\}_{R_s}(U) = -<U|{[\nabla F(U) , \nabla H(U)]}_{R_s}>
\end{eqnarray*}
where $F , H \in C^{\infty}({\cal G}^{*} , R)$. As pointed before, these
Poisson-brackets should describe three dynamical systems given by
\begin{eqnarray*}
{dF\over dt}= \{ H, F\}_{R_{s}}
\end{eqnarray*}

{\bf 2}. Our combined scalar product introduced in sec.2 Eq.(2.28), satisfy the
following properties

\ \ \ \ \ \ \ i) Its conformal weight $\Delta[<< , >>] = 0$

\ \ \ \ \ \ \ ii) The Lie subalgebras ${\Sigma}^{+}_{+}$ , ${\Sigma}^{+}_{0}$
and ${\Sigma}^{+}_{-}$ are dual to ${\Sigma}^{-}_{-}$ ,
${\Sigma}^{-}_{0}$ and ${\Sigma}^{-}_{+}$ with respect to this
combined scalar product in the following sens:
\begin{eqnarray*}
<<{\Sigma}^{+}_{+} , {\Sigma}^{-}_{-}>> &=&  <<R_{s}{\Sigma}^{+}_{+} ,
{\Sigma}^{-}_{-}>>  =  -<<{\Sigma}^{+}_{+} , R_{s}{\Sigma}^{-}_{-}>>\nonumber\\
<<{\Sigma}^{+}_{0} , {\Sigma}^{-}_{0}>> &=& <<R_{s}{\Sigma}^{+}_{0} ,
{\Sigma}^{-}_{0}>>  =  -<<{\Sigma}^{+}_{0} , R_{s}{\Sigma}^{-}_{0}>>\nonumber\\
<<{\Sigma}^{+}_{-} , {\Sigma}^{-}_{+}>>  &=& <<R_{s}{\Sigma}^{+}_{-} ,
{\Sigma}^{-}_{+}>>  =  -<<\Sigma^{+}_{-} , R_{s}{\Sigma}^{-}_{+}>>\nonumber\\
\end{eqnarray*}
showing that $R_{s}$ is skew symmetric with respect to the combined
scalar product $<< , >>$.

3. Each ${\cal G}_{+} = \Sigma_{+}^{+}\oplus \Sigma_{+}^{-}$ and ${\cal
G}_{-} = \Sigma_{-}^{+}\oplus \Sigma_{-}^{-}$ are isotropic with respect
to the combined scalar product $<< , >>$. Moreover ${\cal G}_{0} =
\Sigma_{0}^{+}\oplus \Sigma_{0}^{-}$ is orthogonal to ${\cal G}_{\pm}$.

These properties follow naturally from the definition of the product $<<
, >>$. Indeed for any $X, Y \in {\cal G}_{+}$ (or $X, Y \in
{\cal G}_{-}$) we have
\begin{eqnarray*}
<<X , Y>> = 0
\end{eqnarray*}
Remark finally that this result can be also obtained from the skew symmetry of
the $R_{s}$ with respect to $<< , >>$. We deduce also from Eq.(2.28) that
${\cal G}_{0}$ is orthogonal to ${\cal G}_{\pm}$ since
\begin{eqnarray*}
<<X , Y>> = 0
\end{eqnarray*}
for all $X \in{\cal G}_{\pm}$ and $Y \in{\cal G}_{0}$.

\newpage
\section{conclusion}
We have proposed a consistent approach for the AKS construction of
integrable systems. This approach is based on the conformal spin
decomposition of the huge Lie algebra $\cal G$ of pseudo-differential
operators Eq.(3.13). We used this decomposition to derive three additional
Lie brackets associated to three GMCYB equations. The method exposed
here allows, furthermore to draw a consistent classification principle
of integrable systems.
Future works, will focus to apply the present formulation to
KP-integrable hierarchies which proved to be relevant for a variety of
physical problems both in the bosonic and supersymmetric cases.
\section*{Acknowledgements}
   One of the authors M.B.S would like to thank the ICTP for good hospitality.

\end{document}